\documentclass[twocolumn,showpacs,preprintnumbers,superscriptaddress,amsmath,amssymb]{revtex4-1}

\usepackage{epsfig}
\usepackage{graphicx}
\usepackage{dcolumn}
\usepackage{bm}
\usepackage{times}
\usepackage{xcolor}
\begin{document}

\title{Dynamics of social contagions with heterogeneous adoption thresholds: Crossover phenomena in phase transition}

\author{Wei Wang}
\affiliation{Web Sciences Center, University of Electronic
Science and Technology of China, Chengdu 610054, China}
\address{Big data research center, University of Electronic
Science and Technology of China, Chengdu 610054, China}

\author{Ming Tang}  \email{tangminghuang521@hotmail.com}
\affiliation{Web Sciences Center, University of Electronic
Science and Technology of China, Chengdu 610054, China}
\address{Big data research center, University of Electronic
Science and Technology of China, Chengdu 610054, China}

\author{Panpan Shu}
\affiliation{Web Sciences Center, University of Electronic
Science and Technology of China, Chengdu 610054, China}
\address{Big data research center, University of Electronic
Science and Technology of China, Chengdu 610054, China}

\author{Zhen Wang} \email{zhenwang0@gmail.com}
\affiliation{Interdisciplinary Graduate School of Engineering Sciences, Kyushu University, Kasuga-koen, Kasugashi, Fukuoka 816-8580, Japan}

\begin{abstract}
Heterogeneous adoption thresholds exist widely
in social contagions, but were always neglected in previous studies.
We first propose a non-Markovian spreading threshold model with general adoption
threshold distribution. In order to understand the effects of
heterogeneous adoption thresholds quantitatively,
an edge-based compartmental theory is developed for the proposed model.
We use a binary spreading threshold model as a specific example, in which some
individuals have a low adoption threshold (i.e., activists) while the
remaining ones hold a relatively high adoption threshold (i.e.,
bigots), to demonstrate that heterogeneous adoption thresholds
markedly affect
the final adoption size and phase transition. Interestingly, the first-order,
second-order and hybrid phase transitions can be found in the system.
More importantly, there are two different kinds of crossover phenomena in phase
transition for distinct values of bigots' adoption threshold: a change
from first-order or hybrid phase transition to the second-order phase transition.
The theoretical predictions based on the suggested theory
agree very well with the results of numerical simulations.
\end{abstract}



 \maketitle
\section{Introduction}
Social contagions are ubiquitous in human society, and generate new scientific
challenges for network science~\cite{Watts2007,Castellano2009}.
All the researches about sentiments contagion, information spreading and behavior spreading fall into the category of studying social contagions~\cite{Christakis2007,Barrat2008}.
In particular, behavior spreading, as a representative and essential type
of social
contagions, has attracted great attention, both theoretically and experimentally~\cite{Centola2011,Banerjee2013}. Understanding
the spreading mechanisms behind behavior has the potential to not only help us design
better anti-virus strategies~\cite{Pastor-Satorras2014}, but also shed new insights into
the control of social unrest~\cite{Watts2007}. Moreover, different from biological
contagions (such as epidemic spreading)~\cite{Moreno2002,Pastor-Satorras2001,Salathe2011,Yang2015b,Yang2011,
Li2014},
social contagions display one inherent characteristic: social reinforcement
effect~\cite{Pastor-Satorras2014,Porter2014}, which usually plays a vital role in the
final status of contagions.

To examine social reinforcement effect, some successful models have been proposed~\cite{Watts2002,Granovetter1973}, most of which assume that all
individuals have the same adoption threshold to incorporate such an
effect (so-called threshold model). That is to say, each individual adopts the behavior
only when the fraction~\cite{Watts2002} or number~\cite{Granovetter1973} of neighbors
with the adopted state exceeds his adoption
threshold. In this case, the social contagions is a trivial case of Markovian process.
By means of numerical simulations and theoretical analysis, it was found that the
social reinforcement effect can evidently change the phase transitions of   contagion
dynamics. More specifically, the final adoption size first grows continuously and then
decreases discontinuously with the increase of mean degree when
the adoption threshold of all individuals is identical~\cite{Watts2002}.
After this seminal discovery, the role
of threshold model and its various underlying mechanisms, in social contagions have been
intensively explored, including the influence of dynamical parameters (e.g., initial
seeds and threshold sizes~\cite{Gleeson2007,Singh2013}), and topology characteristics
(degree-degree correlations~\cite{Dodds2009}, clustering~\cite{Whitney2007,Gleeson2008}, community structure~\cite{Nematzadeh2014} as well as multiplexity
framework~\cite{Brummitt2012,Yagan2013}), which are the primary factors in determining the
final adoption size and phase transition. Some non-Markovian social contagion models were
also proposed to describe the social reinforcement effect~\cite{Dodds2004,Wang2015,Zheng2013}.
For example,
in a recent research paper~\cite{Wang2015}, where the social reinforcement was derived from the
memory of non-redundant information transmission, it was found that the perturbation of
dynamical or structural parameters makes the dependence of final behavior adoption size on
information transmission probability
change from being discontinuous to being continuous. In
Ref.~\cite{Zheng2013}, Zheng \emph{et al}. further verified that  the role of social
reinforcement in behavior diffusion on regular graphs and online social networks, which is
consistent with early experimental anticipation~\cite{Centola2010}

Recently, researchers found that the widely existed individual heterogeneity dramatically
alters the spreading dynamics. In epidemic spreading, the heterogeneous infectivity
and susceptibility can change the outbreak threshold~\cite{Miller2008,Yang2015}. For
information spreading, the heterogeneous waiting and response time may speed up or
slow down the velocity of information diffusion~\cite{Cui2014,Jo2014}. Statistical
physicists found that individual heterogeneity
can induce the hybrid phase transition~\cite{Wu2014,Hu2011}, which mixes the traditional
first-order and second-order transitions, in \emph{k}-core percolation~\cite{Cellai2011} and bootstrap percolation~\cite{Gleeson2007,Baxter2011,Lee2014}.
In practical  behavior spreading, individuals usually show
different wills to mimic the behavior, which means that each agent
owns his own adoption threshold~\cite{Karsai2014}. Some
individuals with low adoption threshold show strong wills to adopt the behavior and act
as \emph{activists}. Nevertheless, others with high adoption
threshold need to capture more behavioral information before imitation, and
they often act as \emph{bigots}. With regard to the difference of adoption threshold, it may
be closely related  with personal interests, education background, or other
personality and social factors~\cite{Watts2007}. For example, the well-educated population is more likely to adopt the
high-tech products  than that
among the populations who lacks the basic education. Similarly, students are more  likely to
adopt an interesting computer game than   housewives.

Unfortunately, there is still absence of systematical understanding about the role
of heterogeneous adoption thresholds in social contagions. Aiming to resolve this  issue,
we will explore  how heterogeneous adoption thresholds affect the final adoption size and phase
transition of social contagions based on a so-called binary spreading threshold model,
which is a non-Markovian process. Meanwhile,  an edge-based compartmental theory is developed for quantitative validation. Interestingly,
it is found that heterogeneous adoption thresholds significantly  affect the final adoption size, and
generate the hierarchical characteristic of adopting behavior: activists first adopt the given behavior themselves, and
then stimulate bigots to follow this behavior. Moreover, it is worth noting that such a
heterogeneous threshold model results in the existence of first-order, seconde-order
and hybrid phase transitions. More importantly, there are  two different kinds of crossover
phenomena in phase transition for distinct values of the bigots' adoption threshold:
a change from first-order or hybrid phase transitions to the second-order phase
transition.  In what follows, we will first
describe the heterogeneous social contagion model in complex networks, followed by the description of
edge-based compartmental theory, and then represent the simulation and analysis results. Finally,
we will draw  our conclusions.

\section{Social contagion model with heterogeneous adoption thresholds} \label{Model}

Behavior spreading on complex networks is considered with $N$ nodes and
the degree distribution $P(k)$. As the interaction networks, we use the configuration model~\cite{Catanzaro2005} to avoid the additional
influence of degree-degree correlations.
Nodes in the network represent individuals and edges between nodes stand for the contacts with
which behavioral information transmission may occur. For each individual,
a static behavioral adoption threshold is assigned according to a specific distribution function
$F(T)$, which is independent of network topology. The larger value of $T$ means that an
individual needs to capture more behavioral information from his neighbors
before adopting  the behavior.

With regard to behavior spreading dynamics, we generalize the spreading threshold model with
social reinforcement derived from memory of non-redundant information transmission
characters~\cite{Wang2015,Wang2015b}.  In this model, each individual
falls into one of the three
states: susceptible (S), adopted (A) and recovered (R) (namely, susceptible-adopted-recovered, SAR model). In the susceptible state, an individual does not adopt the behavior. In the
adopted state, an individual adopts the behavior and tries to transmit the behavioral
information to his neighbors. In the recovered state, an individual loses interest
in the behavior and will not transmit the behavioral information further. Initially,
a vanishingly small fraction of individuals $\rho_0$  are chosen as
seeds (adopters)
at random, while the others are fixed in the susceptible state.

At each time step, each  adopted individual $v$ tries to diffuse the behavioral information
to every susceptible neighbor with probability $\lambda$. In particular, once the
information is transmitted through an edge successfully, it will never be transmitted again, i.e., only non-redundant information transmission is allowed.
If the susceptible neighbor $u$ of $v$ is successfully informed, his
cumulative pieces of information $m$ add $1$ (i.e., $m=m+1$).
Subsequently, individual $u$ compares the new value of $m$  with his adoption threshold $T_u$,
and becomes an adopter once $m\geq T_u$. Obviously, whether an individual adopts the behavior
is determined by the cumulative pieces of behavioral information he ever received from
distinct neighbors. Thus, the non-Markovian effect is induced in the behavior spreading
dynamics. After information transmission,
individual $v$ may lose interest in the behavior with probability
$\gamma$ and then moves into the recovered state. Individuals falling into
the recovered state will stop from participating in the further
behavioral information spreading, and the spreading dynamics
terminate when
all adopted individuals become recovered.

\section{Edge-based compartmental theory} \label{theory}

In order to describe the  strong dynamical correlations among the states of neighbors in
heterogeneous
social contagion model, an edge-based compartmental approach is established herein, which is inspired by
Refs.~\cite{Miller2011,Miller2013,Wang2014}. Correspondingly, the notations $S(t)$, $A(t)$ and $R(t)$ respectively represent the fraction of individuals in the susceptible, adopted, and recovered states at time step $t$.

\subsection{General adoption threshold distribution} \label{general_theory}

Individual $u$ is set to be in the cavity state, which means that he can
receive behavioral information from neighbors but not transmit
behavioral information to his neighbors~\cite{Karra2010}. Define
$\theta(t)$ as the probability that individual $v$
has not transmitted the behavioral information to individual $u$ along a randomly chosen edge by time $t$. Thus,
the probability that an individual $u$ with degree $k$
has received $m$ pieces of behavioral information from distinct neighbors by time $t$ can be expressed as
\begin{equation} \label{active_a_k}
\phi_m(k,t)=\binom{k}{m}[\theta(t)]^{k-m}[1-\theta(t)]^m.
\end{equation}
Individual $u$ in the susceptible state implies that the cumulative pieces of behavioral
information $m$ he received are  still less than his adoption
threshold $T_u$. According to the social contagion model
in Sec.~\ref{Model}, the adoption threshold and degree are independent. Considering all
possible values of $m$ and $T_u$, it can be obtained that
the probability of individual $u$ with degree $k$ being susceptible is
\begin{equation} \label{S_K_T}
s(k,t)=\sum_{T_u}F(T_u)\sum_{m=0}^{T_u-1}\phi_m(k,t).
\end{equation}
Combing the degree distribution of a network, the fraction of susceptible individuals
at time step $t$ is
\begin{equation} \label{S_T}
S(t)=\sum_{k}P(k)s(k,t).
\end{equation}
It is obvious that $S(t)$ can be figured out after $\theta(t)$
is known.

As a neighbor of individual $u$ may be in one of susceptible, adopted
or recovered state, $\theta(t)$ can be divided into three cases as
\begin{equation} \label{theta}
\theta(t)=\xi_S(t)+\xi_A(t)+\xi_R(t).
\end{equation}
And $\xi_S(t)$ [$\xi_A(t)$ or $\xi_R(t)$] denotes that the susceptible (adopted or recovered)
neighbor $v$ of $u$ has not transmitted the behavioral information to individual $u$ up to
time $t$.

Then, let's explore the above three terms. If individual $v$ with degree $k^\prime$
is susceptible initially, he can not transmit the behavioral information to $u$, but only receive it from other $k^\prime-1$
neighbors since $u$ in a cavity state. Thus, the probability that individual $v$
has received $m$ pieces of behavioral information by time $t$ is
\begin{equation} \label{tao_S}
\tau_m(k^\prime,t)=\binom{k^\prime-1}{m}
[\theta(t)]^{k^\prime-m-1}[1-\theta(t)]^m.
\end{equation}
Taking all possible values of $m$ and $T_v$ into consideration, the probability of
individual $v$ remaining in susceptible state becomes [similar to Eq.~(\ref{S_K_T})]
\begin{equation} \label{neighbour_S}
\Theta(k^\prime,t)=\sum_{T_v}F(T_v)\sum_{m=0}^{T_v-1}\tau_m(k^\prime,t).
\end{equation}
The probability of an edge connecting an individual with degree
$k^\prime$ is $k^\prime P(k^\prime)/\langle k\rangle$ for uncorrelated networks, where
$\langle k\rangle$
is the mean degree. Thus, it can be obtained that
\begin{equation} \label{xi_S}
\xi_S(t)=\frac{\sum_{k^\prime}k^\prime P(k^\prime)\Theta(k^\prime,t)}{\langle k\rangle}.
\end{equation}

Subsequently, we turn to the expressions of $\xi_A(t)$ and $\xi_R(t)$. Once the behavioral
information is transmitted through an edge with probability $\lambda$, the edge will no longer
satisfy the definition of $\theta(t)$. Thus, the rate of
flow from $\theta(t)$ to $1-\theta(t)$ is $\lambda\xi_A(t)$, which can be expressed as
\begin{equation} \label{d_theta}
\frac{d\theta(t)}{dt}=-\lambda\xi_A(t).
\end{equation}
For the growing of $\xi_R(t)$, two events conditions must be met
simultaneously. At time $t$,
the behavioral information does not transmit through an edge with probability $1-\lambda$, and the adopted individual enters recovered state
with probability $\gamma$. Then,
\begin{equation} \label{xi_R}
\frac{d\xi_R(t)}{dt}=\gamma(1-\lambda)\xi_A(t).
\end{equation}
Combining Eqs.~(\ref{d_theta}) and (\ref{xi_R}), one has that
\begin{equation} \label{xi_R_2}
\xi_R(t)=\frac{\gamma[1-\theta(t)](1-\lambda)}{\lambda}.
\end{equation}
Inserting Eqs.~(\ref{xi_S}) and (\ref{xi_R_2}) into Eq.~(\ref{theta}),
the following expression can be obtained that
\begin{equation} \label{xi_I_2}
\xi_A(t)=\theta(t)-\frac{\sum_{k^{\prime}} k^\prime P(k^\prime)\Theta(k^\prime,t)}{\langle k\rangle}
-\frac{\gamma[1-\theta(t)](1-\lambda)}{\lambda}.
\end{equation}
Substituting Eq.~(\ref{xi_I_2}) into Eq.~(\ref{d_theta}), we get
the time evolution of $\theta(t)$ in details
\begin{equation} \label{d_theta_2}
\frac{d\theta(t)}{dt}=-\lambda[\theta(t)-\frac{\sum_{k^{\prime}} k^{\prime}P(k^{\prime})\Theta(k^{\prime},t)}
{\langle k\rangle}]
+\gamma[1-\theta(t)](1-\lambda).
\end{equation}

Susceptible individuals
move into the adopted states once they adopt the behavior, meanwhile the adopted individuals
may lose interest in the behavior and become recovered. Thus, we can easily get the
evolution of adopted and recovered individuals as
\begin{equation} \label{rho_t}
\frac{dA(t)}{dt}=-\frac{dS(t)}{dt}-\gamma A(t)
\end{equation}
and
\begin{equation} \label{r_T}
\frac{dR(t)}{dt}=\gamma A(t),
\end{equation}
respectively. Eqs.~(\ref{active_a_k})-(\ref{S_T}) and (\ref{d_theta_2})-(\ref{r_T})
give us a complete and
general description of heterogeneous social contagions, from which the
fraction in each state at arbitrary time step can be calculated. When $t\rightarrow\infty$,
we can get the final adoption size $R(\infty)$.

\subsection{Binominal adoption threshold distribution} \label{binary_theory}
In this subsection, we pay attention to the behavior adoption threshold with a binominal
distribution $F(T)$. More specifically, a fraction of individuals $p$ have a
relatively low adoption threshold $T_a$, whereas the remaining individuals have a high adoption threshold $T_b$. $F(T)$ can be expressed as
\begin{equation} \label{Binominal}
 F(T)=\left\{
\begin{array}{rcl}
{T_a,}       &   {\mathrm{with~probability}~p,}\\
{T_b,}     &    {\mathrm{with~probability}~1-p.}
\end{array} \right.
\end{equation}
For simplicity, the values of adoption thresholds  are defined as $T_a=1$ and $T_b\geq1$.
Individuals with low adoption threshold $T_a$ are considered as   \emph{activists}, while those  with
high adoption threshold $T_b$ are regarded as \emph{bigots}. We herein
name this kind of social contagion model as \emph{binary spreading threshold model}.
And the edge-based compartmental theory is utilized  to analyze the
binary spreading threshold model by substituting Eq.~(\ref{Binominal}) into various
equations that give the solutions of $S(t)$, $A(t)$ and $R(t)$. Particularly, we rewrite Eqs.~(\ref{S_K_T})
and (\ref{neighbour_S}) as
\begin{equation} \label{S_K_T_S}
s(k,t)=p\theta(t)^k+(1-p)\sum_{m=0}^{T_b-1}\phi_m(k,t),
\end{equation}
and
\begin{equation} \label{neighbour_S_S}
\Theta(k^\prime,t)=p\theta(t)^{k^\prime-1}+
(1-p)\sum_{m=0}^{T_b-1}\tau_m(k^\prime,t),
\end{equation}
respectively. At time $t$, the fractions of susceptible individuals in the
activist and bigot populations are given by
\begin{equation} \label{S_K_T_L}
S_l(t)=\sum_{k}P(k)\theta(t)^k,
\end{equation}
and
\begin{equation} \label{S_K_T_H}
S_h(t)=\sum_{k}P(k)\sum_{m=0}^{T_b-1}\phi_m(k,t),
\end{equation}
respectively. Considering the fractions of the activist and bigot populations,
the density of susceptible individuals at time step $t$ can also be written as
\begin{equation} \label{S_K_T_Ano}
S(t)=pS_l(t)+(1-p)S_h(t).
\end{equation}

The effects of heterogeneous adoption threshold on phase transition
is another issue concerned. To analyze the phase transition, we can address
the fixed point (root) of Eq.~(\ref{d_theta_2}) at the steady state (i.e., $t\rightarrow\infty$)
with Eq.~(\ref{neighbour_S_S}). That is the fixed point of
\begin{equation} \label{d_theta_steady_general}
\theta(\infty)=y[\theta(\infty)],
\end{equation}
where
\begin{equation} \label{right}
y[\theta(\infty)]=\frac{\sum_{k^{\prime}} k^{\prime}P(k^{\prime})\Theta(k^{\prime},\infty)}
{\langle k\rangle}
 +\frac{\gamma[1-\theta(\infty)](1-\lambda)}{\lambda}.
\end{equation}
From Fig.~\ref{bifurcate}(a), it can be seen that the number of nontrivial roots is either $0$, $1$
or $3$ when $T_b\geq3$. With $T_b=2$, the number of nontrivial roots is either
$0$, $1$ or $2$ [see Fig.~\ref{bifurcate}(b)].

\subsubsection{Case of \textbf{$T_b\geq3$}} \label{T_b_3}

\begin{figure}
\begin{center}
\epsfig{file=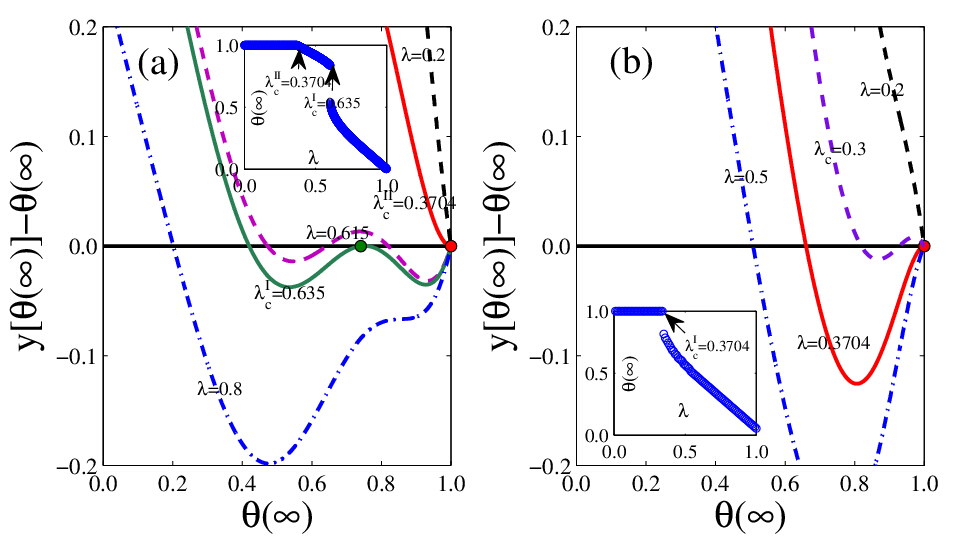,width=1\linewidth}
\caption{(Color online) Illustration of graphical solutions of Eq.~(\ref{d_theta_steady_general}) when  $T_b=4$ (a) and $T_b=2$ (b)
on random regular networks.
In (a) and (b), the black solid lines represent
the horizontal axis and the red
and green dots denote the tangent points.
The inset of (a) and (b) shows the physically meaningful solutions of Eq.~(\ref{d_theta_steady_general}) when $T_b=4$ and $T_b=2$, respectively.
And the arrows indicate critical transmission probabilities.
The fixed points of Eq.~(\ref{d_theta_steady_general}) are the
intersections between the curve and horizontal axis.
Other parameters are defined $p=0.3$ and $\langle k\rangle=10$.}
\label{bifurcate}
\end{center}
\end{figure}

In this subsection, we discuss the case of $T_b\geq3$.
For the given $P(k)$, $p$ and $\gamma$,
Eq.~(\ref{d_theta_steady_general}) has only one trivial solution $\theta(\infty)=1$
when $\lambda$ is small. With the increase of $\lambda$, $\theta(\infty)$ decreases
continuously to a nontrivial solution $\theta(\infty)<1$ first [see example in
Fig.~\ref{bifurcate}(a)], which means that $R(\infty)$ grows continuously first. That
is to say, there is a second-order (continuous) phase transition.
By setting $\theta(\infty)$ and $y[\theta(\infty)]$
tangent at $\theta(\infty)=1$~\cite{Hu2011,Newman2001}, we get the continuous critical information
transmission probability as
\begin{equation}\label{lambda_II_general}
\lambda_c^{II}=\frac{\gamma\langle k\rangle}{p(\langle k^2\rangle-\langle k\rangle)-(1-\gamma)\langle k\rangle},
\end{equation}
where $\langle k\rangle$ and $\langle k^2\rangle$ are the first and second moments of degree
distribution, respectively.
The critical value $\lambda_c^{II}$ separates the local behavior adoption (i.e., the behavior
can be adopted by a vanishingly small fraction of individuals) from the global behavior
adoption (i.e., the behavior can be adopted by a finite fraction of individuals).
From Eq.~(\ref{lambda_II_general}), it is discovered that the occurrence of global behavior adoption
is determined by the network topology (i.e., degree distribution), the fraction of
activists $p$ and the recovery probability
$\gamma$. The global behavior adoption is more likely outbreak (i.e., a lower
$\lambda_c^{II}$) for scale-free networks with divergent second
moment degree distribution [i.e., $\langle k^2\rangle\rightarrow\infty$].
Increasing the value of $p$ can facilitate the global
behavior adoption (i.e., a lower $\lambda_c^{II}$). When $p=1$,
Eq.~(\ref{lambda_II_general}) returns to the case of epidemic outbreak
threshold~\cite{Wang2014}.

Fixing all the parameters except $p$,
similar to Eq.~(\ref{lambda_II_general}), we get the continuous critical fraction of activists as
\begin{equation}\label{p_II_general}
p_c^{II}=\frac{\gamma\langle k\rangle}{\lambda(\langle k^2\rangle-\langle k\rangle)}+
\frac{(1-\gamma)\langle k\rangle}{\langle k^2\rangle-\langle k\rangle}.
\end{equation}
From Eq.~(\ref{p_II_general}), it can be  known that
an enough fraction of
activists are necessary for triggering the global behavior adoption,
and $p_c^{II}$ decreases with the increase of network heterogeneity,
$p$ and $\lambda$.
By setting $\gamma=0$ in Eq.~(\ref{p_II_general}), another critical
proportion of activists $p_c^{\star}$ can be obtained, below which any values of $\lambda$
can not trigger the global behavior adoption, which is given by
\begin{equation}\label{p_I_general}
p_c^{\star}=\frac{\langle k\rangle}{\langle k^2\rangle-\langle k\rangle}.
\end{equation}
It is worth noting that Eq.~(\ref{p_I_general}) is the same with network's percolation
condition~\cite{Newman2001}, which means that the global behavior adoption is possible only
when the activists percolate the entire network (i.e., activists can form a finite connected
cluster).

As shown in Fig.~\ref{bifurcate}(a), three nontrivial
roots of Eq.~(\ref{d_theta_steady_general}) occur when $\lambda$ is large enough
[see Fig.~\ref{bifurcate}(a) for $\lambda=0.615$]. This phenomenon is caused by the bigots, since Eq.~(\ref{d_theta_steady_general}) has at most one nontrivial root
when the bigots is absent~\cite{Wang2015}. In this case, the meaningful
solution will be given by the largest stable root (since only this value can be
achieved physically). For $\lambda=\lambda_c^I=0.635$, the tangent point is
the solution. For $\lambda>\lambda_c^I$, the meaningful solution is the only
stable fixed point. The meaning solution of Eq.~(\ref{d_theta_steady_general}) changes abruptly
to a small value from a relatively large value at $\lambda_c^I$ [see the insert
Fig.~\ref{bifurcate}(a)], and resulting in a discontinuous growth of $R(\infty)$. Based on bifurcation theory~\cite{Strogatz1994},  the discontinuous
critical information transmission probability is gained as follows
\begin{equation} \label{First_Order_Exp}
\lambda_c^I=\frac{\gamma}{\Delta+\gamma-1},
\end{equation}
where
\begin{displaymath}
\Delta=\frac{\sum_{k^{\prime}} k^{\prime}P(k^{\prime})
\frac{d\Theta(k^{\prime},\infty)}{d\theta(\infty)}|_{\theta_s(\infty)}}
{\langle k\rangle},
\end{displaymath}
and $\theta_s(\infty)$ is the fixed point of Eq.~(\ref{d_theta_steady_general}).
Combining Eqs.~(\ref{tao_S}) and (\ref{neighbour_S_S}), we get
\begin{equation} \label{d_phi}
\frac{d\Theta(k^\prime,\infty)}{d\theta(t)}=p(k^\prime-1)\theta(\infty)^{k^\prime-2}+
(1-p)\mathcal{X},
\end{equation}
where
\begin{equation}
\begin{split}
\mathcal{X}&=\sum_{m=0}^{T_b-1}\binom{k^\prime-1}{ m}\{(k^\prime-m-1)\theta(\infty)^{k^\prime-m-2}
[1-\theta(\infty)]^m\\
&-
m\theta(\infty)^{k^\prime-m-1}[1-\theta(\infty)]^{m-1}\}.
\end{split}
\end{equation}
Using the analytical method similar to Eq.~(\ref{First_Order_Exp}), the discontinuous
critical fraction of activists can be expressed as
\begin{equation}\label{p_II_dis}
p_c^{I}=\frac{\lambda(1-\mathcal{Y})+\gamma(1-\lambda)}{\lambda(\mathcal{Z}-\mathcal{Y})},
\end{equation}
where $$
\mathcal{Y}=\frac{\sum_{k^\prime}k^\prime P(k^\prime)\mathcal{X}}{\langle k\rangle}
$$
and
$$\mathcal{Z}=\frac{\sum_{k^\prime}k^\prime(k^\prime-1) P(k^\prime)\theta^{k^\prime-2}}{\langle k\rangle}.$$

From the above analysis, we find that $R(\infty)$ versus $\lambda$ or $p$   first
grows continuously and then follows a discontinuous fashion. And the continuous and discontinuous growthes
of $R(\infty)$ are caused by the activists and bigots, respectively,
which can be regarded as hybrid phase transition, mixing
the traditional first-order and
second-order transitions, from the perspective of statistical
physics~\cite{Dorogovtsev2008}. Note that the hybrid phase transition
can change to a second-order phase transition, and
does not exist under any conditions. Numerically solving Eqs.~(\ref{d_theta_steady_general}),
(\ref{First_Order_Exp}), and the following equation
\begin{equation} \label{First_Order_Condition}
\frac{d^2g[\theta(\infty)]}{d\theta^2(\infty)}=0,
\end{equation}
we can learn the condition under which the hybrid phase transition disappears.

\subsubsection{Case of $T_b=2$} \label{T_b_2}
We study the special case of $T_b=2$ in this subsection. As shown
in Fig.~\ref{bifurcate}(b), for any $\lambda$, $y[\theta(\infty)]$
can only tangent to $\theta(\infty)$ when  $\theta(\infty)=1$, and can not
tangent to other $\theta(\infty)<1$. In addition,
Eq.~(\ref{d_theta_steady_general}) has $1$ or $2$ nontrivial
fixed points. These phenomena
means that the meaningful solution of Eq.~(\ref{d_theta_steady_general})
jumps to another one at the critical information transmission
probability [see the inset of Fig.~\ref{bifurcate}(b)]. As a result,
$R(\infty)$ increases discontinuously versus $\lambda$.
The critical information transmission
probability can be acquired in the similar way as when
$T_b=3$. The values of $\lambda_c^{II}$ and
$\lambda_c^I$ can be obtained from Eq.~(\ref{lambda_II_general}),
since $y[\theta(\infty)]$
can only tangent to $\theta(\infty)$ when  $\theta(\infty)=1$.
Numerical solving Eqs.~(\ref{d_theta_steady_general}), (\ref{First_Order_Exp})
and (\ref{First_Order_Condition}), we can get the condition under which
the first-order phase transition changes to a second-order phase transition.

\section{Numerical verification}
In the study, extensive simulations are conducted for
on binary spreading threshold model on uncorrelated networks.
Unless otherwise specified, the network size, mean degree and recovery
probability are of $N=10,000$, $\langle k\rangle=10$ and $\gamma=1.0$, respectively.
At least $2\times10^3$ independent dynamical
realizations on a fixed network are used to calculate the pertinent average
values, which are further averaged over $100$ network realizations.

The \emph{relative variance} $v_R$ is applied to numerically determine the size-dependent
critical values, such as, $\lambda_c^I$, $\lambda_c^{II}$, $p_c^I$ and $p_c^{II}$.
The relative variance of $R(\infty)$ is defined as
\begin{equation} \label{relative_V}
v_R=\frac{\langle R(\infty)-\langle R(\infty)\rangle\rangle^2}{\langle R(\infty)\rangle^2},
\end{equation}
where $\langle\cdots\rangle$ denotes ensemble averaging.
The value of $v_R$ exhibits peaks at the phase transition, which announce the phase
transition~\cite{Chen2014}. We determine the critical value $\omega_c$, which represents
$\lambda_c^I$, $\lambda_c^{II}$, $p_c^I$ or $p_c^{II}$, as the value
of $\lambda$ or $p$ when the relative variance reaches its maximum, and
\begin{equation} \label{num_cri}
\omega_c=\mathrm{arg}\{\mathrm{max}~v_R\}.
\end{equation}
Note that Eq.~(\ref{num_cri}) is not the only possible way to compute the critical values,
other methods can be used to determine $\omega_c$, such as, susceptibility~\cite{Radicchi2014,Ferreira2012}
and variability~\cite{Shu2014}.

\begin{figure}
\begin{center}
\epsfig{file=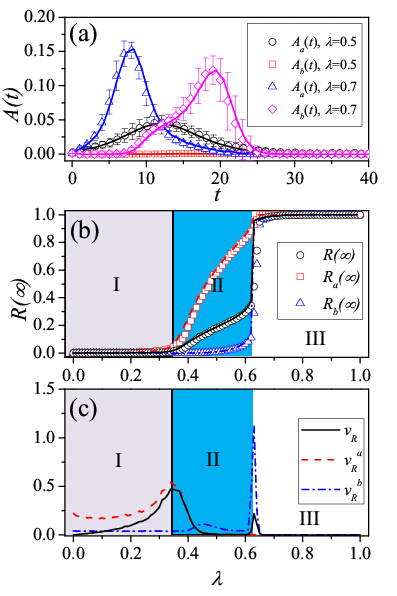,width=0.7\linewidth}
\caption{(Color online) Binary spreading threshold model on random regular networks.
(a) Average density of adopted individuals $A_a(t)$ and $A_b(t)$ versus time $t$
in the activist and bigot populations with  different information transmission
 probability $\lambda$. The error bars
indicate the standard deviations.
(b) Final adoption size $R(\infty)$ and (c) relative variance $v_R$ versus $\lambda$
for full, activist and bigot populations. In figures (a) and (b), symbols represent
the simulated results, and the lines are the corresponding theoretical
predictions from Eqs.~(\ref{d_theta_2})-(\ref{r_T}) and (\ref{S_K_T_S})-(\ref{S_K_T_Ano}). Lines in (c) are the simulation results of $v_R$. In (b) and (c), the vertical lines
are the critical continuous and discontinuous information transmission probabilities,
which are denoted as $\lambda_c^{II}$ and $\lambda_c^I$, respectively. The theoretical
values of $\lambda_c^{II}$ and $\lambda_c^I$ can be gotten from
Eqs.~(\ref{lambda_II_general}) and (\ref{First_Order_Exp}), respectively.
The theoretical (numerical) values of $\lambda_c^{II}$ and $\lambda_c^I$
separate (b) [(c)] into three regions. Region I, only a vanishingly small fraction of individuals can be exposed to adopt the behavior; region II, only a finite
fraction of activists adopt the behavior; region III, a finite fraction of
activists and bigots adopt the behavior. Other parameters are set to be $N=10,000$,
$p=0.3$ and $T_b=4$.}
\label{fig1}
\end{center}
\end{figure}

\subsection{Random regular networks}
To be illustrative, we first focus on random regular networks (RRNs).
Fig.~\ref{fig1}(a) shows the time evolution of the fraction of adopted
individuals $A_a(t)$ and $A_b(t)$ in the activist and bigot populations
with  different behavioral information transmission probabilities $\lambda$.
It is found that a hierarchical character of behavior adoption is caused by heterogeneous adoption thresholds. That is to say,
activists with low $T_a$ first adopt the behavior and then stimulate the bigots with
$T_b$ to adopt the behavior. With a relatively small $\lambda=0.5$, $A_a(t)$ shows a
small peak, and can not stimulate a finite fraction of bigots to adopt the
behavior [$A_b(t)$ does not see an obvious peak]. With a relatively large
$\lambda=0.7$, $A_a(t)$ shows a large peak, and further causing the emergence of a large peak for $A_a(t)$. The heterogeneous adoption threshold distribution
may be used to explain the existence of multimodal in the adoption of serves~\cite{Karsai2014}.
The time evolution can be well predicted by our edge-based compartmental theory.

From Figs.~\ref{fig1}(b) and (c), it can be seen that $R(\infty)$ versus
$\lambda$ shows a hybrid
phase transition, which means that $R(\infty)$ first grows continuously and then follows
a discontinuous fashion. The continuous and discontinuous phase transitions are caused by activists and bigots, respectively.
Similar to Ref.~\cite{Wang2015}, the discontinuous growth of $R(\infty)$ is
caused by those bigots in the subcritical state who adopt the behavior simultaneously.
An individual in such a state has received the behavioral information but has not
yet adopted the behavior, and the number of information pieces from distinct neighbors
is precisely one less than his adoption threshold.
The theoretical (numerical) values of $\lambda_c^{II}$ and
$\lambda_c^I$ separate Figs.~\ref{fig1}(b) [(c)] into three regions.
The theoretical values of $\lambda_c^{II}$ and $\lambda_c^I$ can be gotten from
Eqs.~(\ref{lambda_II_general}) and (\ref{First_Order_Exp}), respectively.
In region I, with $\lambda\leq\lambda_c^{II}$, both activist and bigot populations
adopt the
behavior locally (i.e., only a vanishingly small fraction of individuals adopted the
behavior). In region II, with $\lambda_c^{II}<\lambda\leq\lambda_c^I$,
activists adopt the behavior globally (i.e., a finite fraction of activists adopted
the behavior) and bigots adopt the behavior locally. In region III with $\lambda>\lambda_c^I$,
both the activists and bigots adopt the behavior globally.
The numerical values of $\lambda_c^{II}$ and $\lambda_c^I$ can be obtained by
observing $v_R$ in Fig.~\ref{fig1}(c). For instance, $v_R$ has two peaks,
which means that two phase transitions occur~\cite{Chen2014}. And the first peak
appears at $\lambda_c^{II}$, while the second peak locates at $\lambda_c^I$.
Note that the first (second) peak of $v_R$ shares the same location with the maximal peak
of activist population $v_R^a$ (bigots population $v_R^b$).
In all, $\lambda_c^I$, $\lambda_c^{II}$ and $R(\infty)$ can be well
predicted by our edge-based compartmental theory.

\begin{figure}
\begin{center}
\epsfig{file=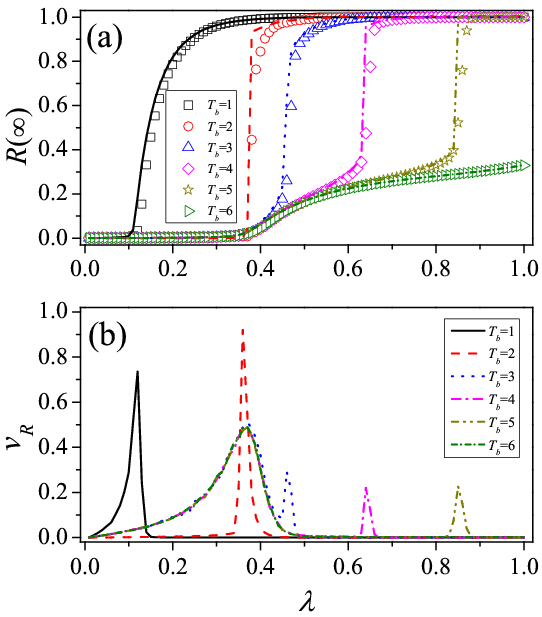,width=1\linewidth}
\caption{(Color online) Effects of bigots' adoption threshold
on binary spreading threshold model. (a) $R(\infty)$ and (b) $v_R$ versus
$\lambda$ with different $T_b$. In figure (a), symbols represent the simulated
results, and the lines are the theoretical predictions from
Eqs.~(\ref{d_theta_2})-(\ref{r_T}) and (\ref{S_K_T_S})-(\ref{S_K_T_Ano}). In figure
(b), lines are the simulation results of $v_R$, and we plot the value of
$v_R/20$ for $T_b=2$. Other parameters are defined as $N=10,000$ and $p=0.3$.}
\label{fig2}
\end{center}
\end{figure}

\begin{figure}
\begin{center}
\epsfig{file=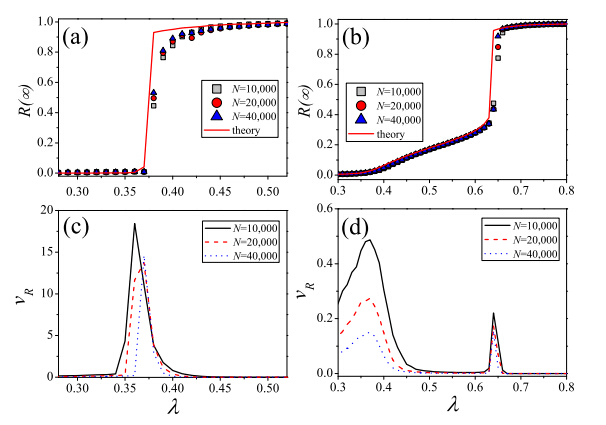,width=1\linewidth}
\caption{(Color online) Finite-size effects on the binary spreading threshold model
for $p=0.3$. (a) $R(\infty)$ and (c) $v_R$ versus $\lambda$ for $T_b=2$ on different network
size $N$. (b) $R(\infty)$ and (d) $v_R$ versus $\lambda$ for $T_b=4$ under different
$N$. The lines in (a) and (b) are the theoretical predictions
from Eqs.~(\ref{d_theta_2})-(\ref{r_T}) and
(\ref{S_K_T_S})-(\ref{S_K_T_Ano}). The lines in (c) and (d)
are the simulation results of $v_R$.}
\label{fig3}
\end{center}
\end{figure}

As shown in Fig.~\ref{fig2}, $R(\infty)$ and the
phase transition are significantly influenced by the adoption
threshold of bigots $T_b$. And $R(\infty)$ decreases with $T_b$,
since a larger value of $T_b$ requires more information to be exposed for bigots.
The phase transition is continuous when $T_b=1$. In this case, the contagion dynamics
is the same with epidemic spreading~\cite{Newman2001}. The phase transition is also
continuous when $T_b\geq6$, since there are not enough activists
to persuade bigots to adopt the behavior simultaneously.
For the case of $T_b=2$, $R(\infty)$ shows a first-order phase transition, because
the bigots are likely to enter subcritical states and adopt the behavior simultaneously.
A hybrid phase transition emerges with other values of $T_b$ (i.e., $2<T_b<6$).
As discussed in Sec.~\ref{binary_theory}, the type of phase transition is verified by bifurcation analysis of Eq.~(\ref{d_theta_steady_general}). In Fig.~\ref{fig2}(b), we verify the phase
transition by studying $v_R$ in simulations. And the simulated values of $\lambda_c^{II}$
and $\lambda_c^I$ are located by studying $v_R$ versus $\lambda$.
For the second-order phase transition, $v_R$ has only one peak [see $T_b=1$ and
$T_b=6$ in Fig.~\ref{fig2}(c)]. Similarly, $v_R$ also has only one peak for the
first-order phase transition [see $T_b=2$ in Fig.~\ref{fig2}(c)].
 For the hybrid phase transition, $v_R$ has two peaks
[see $3\leq T_b\leq5$ in Fig.~\ref{fig2}(c)]. Our theoretical predictions of $R(\infty)$
are well agree with the simulation results, except for the cases
near the critical information transmission probability. The deviations between our predictions and simulation are mainly derived from the finite-size effects of the networks, as shown  in Fig.~\ref{fig3}. The deviations
of $R(\infty)$, $\lambda_c^{I}$ and $\lambda_c^{II}$ between
the simulated and theoretical results decrease with network size $N$.

\begin{figure}
\begin{center}
\epsfig{file=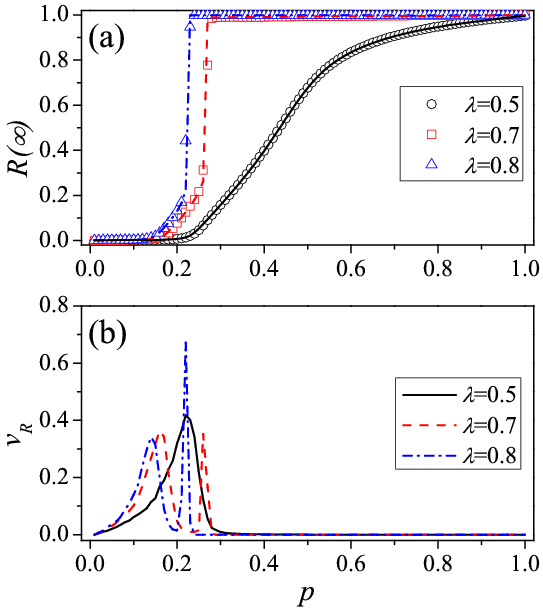,width=1\linewidth}
\caption{(Color online) Effects of the fraction of activists on
binary spreading threshold model. (a) $R(\infty)$ and (b)
$v_R$ as a function of $p$ for different $\lambda$. In figure (a),
symbols represent the simulated results, and the lines are the
theoretical predictions from Eqs.~(\ref{d_theta_2})-(\ref{r_T}) and
(\ref{S_K_T_S})-(\ref{S_K_T_Ano}). Lines in (b) are the simulation results of $v_R$. We set
other parameters to be $N=10,000$ and $T_b=4$.}
\label{fig4}
\end{center}
\end{figure}

Then, we observe $R(\infty)$ versus $p$ for different $\lambda$ in Fig.~\ref{fig4}.
We find that $R(\infty)$ increases with $p$.
By bifurcation analysis of Eq.~(\ref{d_theta_steady_general}) and studying
$v_R$, a continuous growth of $R(\infty)$ is observed with a relatively small $\lambda$
(e.g., $\lambda=0.5$), and the hybrid phase transition occurs with a relatively large
$\lambda$ (e.g., $\lambda=0.7$ and $0.8$). Again, the theoretical
and numerical results agree well.

\begin{figure}
\begin{center}
\epsfig{file=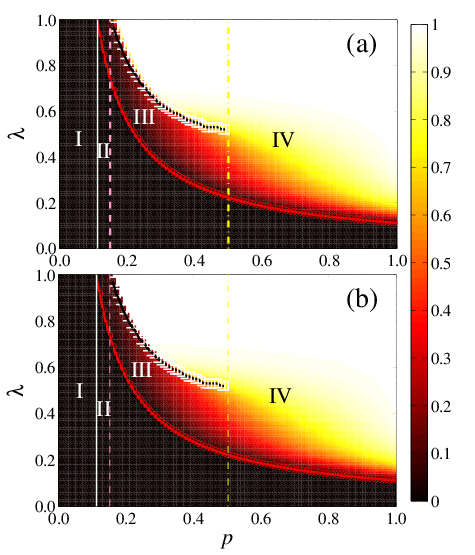,width=1\linewidth}
\caption{(Color online) Dependence of $R(\infty)$ on $p$ and $\lambda$
on random regular networks. Color-coded values $R(\infty)$ are obtained
from numerical simulations
(a) and theoretical solutions (b) in the parameter plane $(p,\lambda)$,
and the theoretical predictions are the solutions of Eqs.~(\ref{d_theta_2})-(\ref{r_T}) and
(\ref{S_K_T_S})-(\ref{S_K_T_Ano}). The plane are divided
into four regions by three vertical lines, of which the
first line can be gotten from
Eq.~(\ref{p_I_general}) and the other two are predicted by solving
Eqs.~(\ref{d_theta_steady_general}), (\ref{First_Order_Exp}) and
(\ref{First_Order_Condition}). In region I, only a vanishingly small
fraction of individuals can be exposed to adopt the behavior (i.e.,
local behavior adoption). Both the regions II and IV show a continuous
phase transition, while region III exhibits a hybrid phase transition.
The red circles (red solid curve) and
white squares (black dished curve) are the continuous and discontinuous
simulated (theoretical) critical information transmission probability,
respectively. Moreover, other parameters defined as $N=10,000$,
$p=0.3$ and $T_b=4$.}
\label{fig5}
\end{center}
\end{figure}

From the above analysis, it can be obtained  that both
$\lambda$ and $p$ markedly affect $R(\infty)$ and phase transition. Thus, we further
investigate $R(\infty)$ and phase transition on parameter plane $(\lambda,p)$ when
$T_b=4$ in Fig.~\ref{fig5}. Obviously, $R(\infty)$ increases
with $\lambda$ and $p$. According to the type of phase transition,
the parameter plane $(\lambda,p)$ is divided into four different regions by
three vertical lines. The first vertical line can be gotten from Eq.~(\ref{p_I_general}),
and the other two can be predicted by solving
Eqs.~(\ref{d_theta_steady_general}), (\ref{First_Order_Exp}) and
(\ref{First_Order_Condition}).
In region I ($p\leq p_c^{\star}=1/9$), there are a few activists,
who can not percolate
the entire population. Thus, no matter what the value of
$\lambda$,  activists can not be made to
adopt the behavior globally. When $p>p_c^\star$, the
global behavior adoption becomes possible, and a crossover
phenomenon, which means that the phase transition changes from being hybrid to being second-order, occurs in the phase transition. Meanwhile, the
local and global behavior adoptions are separated by
the red solid curve (i.e., $\lambda_c^{II}$). In region II ($1/9<p\leq0.15$),
the relatively few
activists lead to the continuous phase transition. In this region,
$R(\infty)$ grows continuously versus $\lambda$ for a given $p$, and
a finite fraction of individuals adopt the behavior above $\lambda_c^{II}$.
With the increase of $p$, in region III ($0.15<p<0.5$), the hybrid phase
transition occurs, i.e., $R(\infty)$ first grows continuously with $\lambda$ and then
follows by a discontinuous pattern. A finite fraction of activists adopt the
behavior above $\lambda_c^{II}$, and further induce the bigots to adopt the
behavior simultaneously above $\lambda_c^I$ [see black curves obtained from Eq.~(\ref{First_Order_Exp})]. In region IV ($p\geq0.5$), half of the neighbors
of bigots are   activists. Once these activists
adopt the behavior, the bigots will gradually adopt the behavior. Thus,
$R(\infty)$ grows continuously and a finite fraction of individuals adopt the
behavior above $\lambda_c^{II}$ (see red curves). Our theoretical
predictions of $\lambda_c^{II}$, $\lambda_c^{I}$ and $R(\infty)$ have a good
agreement with the numerical predictions.

\begin{figure}
\begin{center}
\epsfig{file=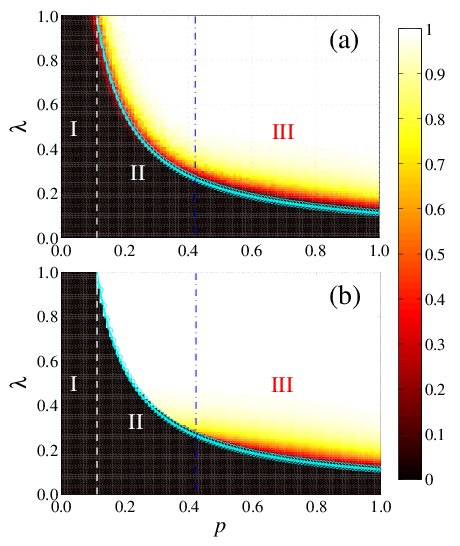,width=1\linewidth}
\caption{(Color online) Dependence of $R(\infty)$
on $p$ and $\lambda$. Color-coded values $R(\infty)$ from numerical simulations
(a) and theoretical solutions (b) in the parameter plane $(p,\lambda)$,
and the theoretical predictions are the solutions of Eqs.~(\ref{d_theta_2})-(\ref{r_T}) and
(\ref{S_K_T_S})-(\ref{S_K_T_Ano}). Two vertical lines separate the
plane into three regions, the former line is predicted by
Eq.~(\ref{p_I_general}) and the latter line is predicted by numerically solving
Eqs.~(\ref{d_theta_steady_general}), (\ref{First_Order_Exp}) and
(\ref{First_Order_Condition}). In region I, only a vanishingly small
fraction of individuals can be exposed to adopt the behavior (i.e.,
local behavior adoption). Region II shows a discontinuous
phase transition, while region III exhibits a continuous phase transition.
The blue circles (blue dished solid curve) are  the
simulated (theoretical) critical information transmission probability,
respectively. We set other parameters as $N=10,000$, $p=0.3$ and $T_b=2$.}
\label{fig6}
\end{center}
\end{figure}

It can be seen in Fig.~\ref{fig2} that $R(\infty)$ increases
discontinuously with $\lambda$ when $T_b=2$, thus $R(\infty)$ and phase
transition on parameter plane $(p,\lambda)$ for $T_b=2$  is explored
in Fig.~\ref{fig6}. And we find another crossover
phenomenon in the phase transition: a change from being first-order to being
second-order. Similar to Fig.~\ref{fig5} is that the plane is divided into three regions: Region I ($p\leq1/9$), the local behavior adoption region, in
which only a vanishingly small fraction of individuals
adopt the behavior; region II ($1/9<p\leq0.42$) shows a first-order
phase transition, where a finite fraction of individuals adopt the behavior
simultaneously above $\lambda_c^I$ (red dished lines); region III ($p>0.42$) exhibits a second-order phase transition, in which $R(\infty)$ increases
continuously versus $\lambda$. The type of
phase transition is verified by bifurcation analysis and studying $v_R$.

\subsection{Heterogeneous networks}
We turn to elucidate the effects of network heterogeneity. To build
the heterogeneous networks, the uncorrelated configuration model
with power-law degree distributions $P(k)\sim k^{-\gamma_D}$ is
adopted, where the mean degree
$\langle k\rangle=10$ and the maximum degree $k_{max}\sim\sqrt{N}$~\cite{Catanzaro2005}.
The network heterogeneity increases with the decrease of
$\gamma_D$.

\begin{figure}
\begin{center}
\epsfig{file=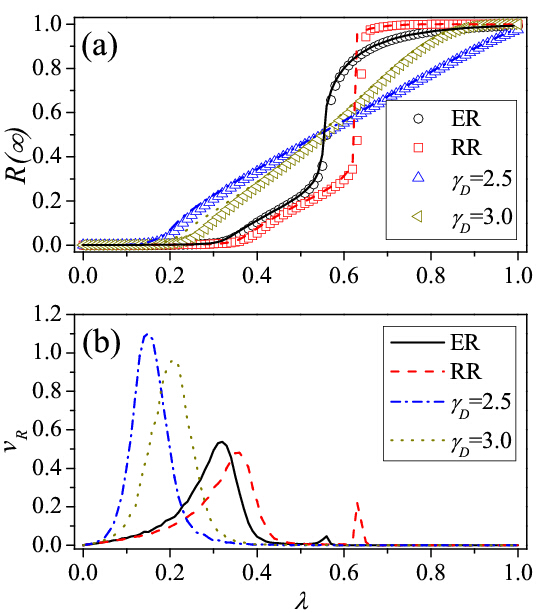,width=1\linewidth}
\caption{(Color online) For $T_b=4$, the effects of network
heterogeneity on binary spreading
threshold model. (a) $R(\infty)$ and (b) $v_R$ versus $\lambda$
on random regular networks, Erd\H{o}s-R\'{e}nyi (ER) networks, and scale-free networks
with degree exponent $\gamma_D=2.5$ and $3.0$. In figure (a),
symbols represent the simulated results, and the lines are
theoretical predictions from Eqs.~(\ref{d_theta_2})-(\ref{r_T}) and (\ref{S_K_T_S})-(\ref{S_K_T_Ano}). Lines in (b) are the simulation results of $v_R$.
We set other parameters to be $N=10,000$ and $p=0.3$.}
\label{fig7}
\end{center}
\end{figure}

In the case of $T_b=4$, we find that the global behavior
adoption more likely to occur
(i.e., lower $\lambda_c^{II}$) in heterogeneous networks, due
to the existence of hubs in heterogeneous networks~\cite{Holme2003},
as shown in Fig.~\ref{fig7}. Meanwhile, in strong heterogeneous networks
a large number of individuals with small degrees are difficult to adopt
the behavior, so $R(\infty)$ is smaller at large $\lambda$.
For example, $R(\infty)$ at $\lambda=0.7$ is obversely smaller on scale-free
networks with $\gamma_D=2.5$ than that on RRNs. By bifurcation analysis of
Eq.~(\ref{d_theta_steady_general}), it is discovered that the
hybrid phase transition
disappears for strong heterogeneous networks (e.g., $\gamma_D=2.5$ and $3.0$
in Fig.~\ref{fig7}). That is to say, network heterogeneity leads to the crossover phenomenon: a change from being hybrid to being second-order.
Moreover, the type of phase transition transition is further verified by
observing $v_R$ in Fig.~\ref{fig7}(b). Evidences in
terms of the quantities $R(\infty)$, $\lambda_c^{I}$ and $\lambda_c^{II}$
support our edge-based compartmental theory.

\begin{figure}
\begin{center}
\epsfig{file=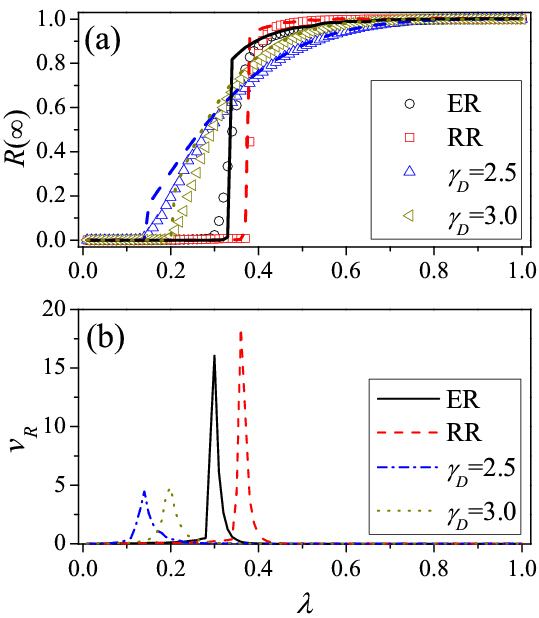,width=1\linewidth}
\caption{(Color online) For $T_b=2$, the effects of network heterogeneity on binary spreading
threshold model. (a) $R(\infty)$ and (b) $v_R$ versus $\lambda$
on random regular networks, Erd\H{o}s-R\'{e}nyi (ER) networks, and scale-free networks
with degree exponent $\gamma_D=2.5$ and $3.0$. In figure (a),
symbols represent the simulated results, and the lines are the
theoretical predictions from Eqs.~(\ref{d_theta_2})-(\ref{r_T}) and (\ref{S_K_T_S})-(\ref{S_K_T_Ano}). Lines in (b) are the simulation results of $v_R$. And other parameters are set to be $N=10,000$ and $p=0.3$.}
\label{fig8}
\end{center}
\end{figure}

For the case of $T_b=2$ (see Fig.~\ref{fig8}), we find the similar phenomenon
of $R(\infty)$: $R(\infty)$ increases (decreases) with network
heterogeneity for small (large) $\lambda$. However, the system has the first-order phase transition, which denotes that network heterogeneity does not alter the phase transition. Based on the bifurcation theory and the
study of $v_R$, the phase transition  is verified
[see Fig.~\ref{fig8}(b)].

\section{Conclusions}
Understanding social contagion dynamics in human populations is extremely challenging. In practical behavior spreading, individuals usually display
different criterions (wills) to adopt the behavior. That is to say, the heterogeneity
of adoption thresholds do indeed exist, but its effects
on social contagions have not been verified straightforward.
To fill this gap, we proposed a non-Markovian behavior spreading
model, in which individuals have distinct adoption thresholds, to explore how
heterogeneous adoption thresholds affect the final adoption size and phase
transition. An edge-based compartmental theory is developed to
quantificationally describe this model, and this suggested theory
is verified by a large number of simulations.
In the paper, we mainly focused on the so-called binary spreading
threshold model, in which a fraction of individuals $p$ have
the adoption threshold $T_a=1$
and are acted as activists, and the remaining ones have a higher adoption threshold
$T_b$ and are regarded as bigots.

We first studied the spreading dynamics on random regular
networks. And it is found that heterogeneous adoption thresholds markedly affect
$R(\infty)$, and induce a hierarchical character in behavior adoption.
In other words, activists first adopt the behavior and then stimulate bigots to
adopt the behavior. All the first-order, second-order
and hybrid phase transitions are found to be in existence
in the system. For the case of $T_b=1$, the traditional
second-order phase transition can be found. For $T_b\geq3$, the hybrid phase
transition mixed first and second order, i.e., $R(\infty)$ versus $\lambda$ first
grows continuously and then follows by a discontinuous pattern,
also occurs. More specifically, the continuous and
discontinuous growth of $R(\infty)$ are caused by the activists and
bigots, respectively. There is a crossover phenomenon between the
hybrid and second-order phase transitions by varying $p$.
When $T_b=2$, the system only exhibits the first-order or
second-order phase transitions. Interestingly, there
is another crossover phenomenon: by varying $p$ the phase
transition changes from being first-order to being second-order.

Finally, we found that network heterogeneity markedly affects $R(\infty)$
and phase transition due to the existence of hub individuals.
For the case of $T_b\geq3$, strong network heterogeneity causes the
phase transition to change from being hybrid to being second-order.
For $T_b=2$, network heterogeneity does not alter the type
of phase transition.

The main contribution of our work lies in providing a
qualitative and  quantitative view on the influence of heterogeneous
adoption threshold. Meanwhile, our research results also enrich the phase transition phenomenon.
The developed theory can be
generalized to behavior spreading with general adoption threshold
distribution, and offer some new inspirations for other similar
spreading dynamics, such as epidemic spreading and cascading.
However, some fascinating and hopeful challenges still remain.
For example, what will happen if the adoption thresholds are
correlated with their degrees? How to extract more realistic
behavior spreading mechanisms from real data?

\section*{Acknowledgements}
This work was partially supported by the National Natural Science
Foundation of China under Grants Nos.~11105025 and 11575041,
the Program of Outstanding Ph. D. Candidate in Academic Research by
UESTC under Grand No.~YXBSZC20131065.


\section*{References}
\begin{thebibliography}{100}
\bibitem{Watts2007}
Watts D  J and Dodds P S 2007
Journal of Consumer Research, \textbf{34} 441.

\bibitem{Castellano2009}
Castellano C, Fortunato S and Fortunato S 2009
Rev. Mod. Phys. \textbf{81} 0034.

\bibitem{Christakis2007}
Christakis N A and Fowler J H 2007 N. Engl. J. Med. \textbf{357} 370.

\bibitem{Barrat2008}
Barrat A, Barth\'{e}lemy M, and Vespignani A 2007 \emph{Dynamical
Processes on Complex Networks} (Cambridge: Cambridge University Press).

\bibitem{Centola2011}
Centola D 2011 Science \textbf{334} 1269.

\bibitem{Banerjee2013}
Banerjee A, Chandrasekhar A G, Duflo E and
Jackson M O 2013 Science \textbf{341} 363.

\bibitem{Pastor-Satorras2014}
Pastor-Satorras R, Castellano C,
Mieghem P V and Vespignani A 2014 arXiv:1408.2701v1.

\bibitem{Moreno2002}
Moreno Y, Pastor-Satorras R and Vespignani A
2002 Eur. Phys. J. B \textbf{26} 521.

\bibitem{Pastor-Satorras2001}
Pastor-Satorras R and Vespignani A 2001
Phys. Rev. Lett. \textbf{86} 3200.

\bibitem{Salathe2011}
Salath\'{e} M and Khandelwal S 2011 PLOS Comput. Biol. \textbf{7} e1002199.

\bibitem{Yang2015b}
Yang H X, Tang M, and Lai Y C
2015 Phys. Rev. E \textbf{91} 062817.

\bibitem{Yang2011}
Yang H X , Wang W X, Lai Y C, Xie Y B, and Wang B H
2011 Phys. Rev. E \textbf{84} 045101(R).

\bibitem{Li2014}
Li K, Fu X, Small M, and Zhu G
2014 Chaos \textbf{24} 043124.

\bibitem{Porter2014}
Porter M A and Gleeson J P 2014
arXiv:1403.7663v1.

\bibitem{Watts2002}
Watts D J 2002 Proc. Natl. Acad. Sci. \textbf{99} 5766.

\bibitem{Granovetter1973}
Granovetter M 1973 Am. J. Sociol. \textbf{78}
1360.

\bibitem{Gleeson2007}
Gleeson J P and Cahalane D J 2007 Phys. Rev. E \textbf{75} 056103.

\bibitem{Singh2013}
Singh P, Sreenivasan S, Szymanski B K and Korniss G
2013 Sci. Rep. \textbf{3} 2330.

\bibitem{Dodds2009}
Dodds P S and Payne J L 2009 Phys. Rev. E \textbf{79} 066115.

\bibitem{Whitney2007}
Whitney D E 2010 Phys. Rev. E \textbf{82} 066110.

\bibitem{Gleeson2008}
Gleeson J P 2008 Phys. Rev. E \textbf{77} 046117.

\bibitem{Nematzadeh2014}
Nematzadeh A, Ferrara E, Flammini A
and Ahn Y Y 2014 Phys. Rev. Lett. \textbf{113} 088701.

\bibitem{Brummitt2012}
Brummitt C D, Lee K -M and Goh K-I 2012 Phys. Rev. E \textbf{85}
045102(R).

\bibitem{Yagan2013}
Ya\v{g}an O and Gligor V 2013 Phys. Rev. E \textbf{86} 036103.

\bibitem{Dodds2004}
Dodds P S and Watts D J 2004. Phy. Rev. Lett. \textbf{92}
218701.

\bibitem{Wang2015}
Wang W, Tang M, Zhang H-F and Lai Y-C 2015
Phys. Rev. E \textbf{92} 012820.

\bibitem{Zheng2013}
Zheng M, L\"{u} L and Zhao M 2013 Phys. Rev. E \textbf{88} 012818.

\bibitem{Centola2010}
Centola D 2010 Science \textbf{329} 1194.

\bibitem{Miller2008}
Miller J C 2007 Phys. Rev. E \textbf{76} 010101.

\bibitem{Yang2015}
Yang H, Tang M and Gross T 2015 Sci. Rep. \textbf{5} 13122.

\bibitem{Cui2014}
Cui A-X, Wang W, Tang M, Fu Y,
Liang X and Do Y 2014 Chaos \textbf{24}
033113.

\bibitem{Jo2014}
Jo H H, Perotti J I, Kaski K and Kert\'{e}sz J 2014
Phys. Rev. X \textbf{4} 011041.


\bibitem{Wu2014}
Wu C, Ji S, Zhang R, Chen L, Chen J, Li X and Hu Y 2014 Europhys. Lett. \textbf{107}, 48001.

\bibitem{Hu2011}
Hu Y, Ksherim B, Cohen R and Havlin S 2011
Phys. Rev. E \textbf{84}, 066116.

\bibitem{Cellai2011}
Cellai D, Lawlor A, Dawson K A and Gleeson J P 2011
Phys. Rev. Lett. \textbf{107} 175703.

\bibitem{Baxter2011}
Baxter G J, Dorogovtsev S N, Goltsev A V and Mendes J F F
2011 Phys. Rev. E \textbf{83} 051134.

\bibitem{Lee2014}
Lee K-M, Brummitt C D and Goh K-I 2014 Phys. Rev. E \textbf{90} 062816.

\bibitem{Karsai2014}
Karsai M, I\~{n}iguez G, Kaski K and Kert\'{e}sz J 2014
J. R. Soc. Interface \textbf{11} 101.

\bibitem{Catanzaro2005}
Catanzaro M, Bogu\~{n}\'{a} M and Pastor-Satorras R 2005
Phys. Rev. E \textbf{71} 027103.

\bibitem{Wang2015b}
Wang W, Shu P-P, Zhu Y-X, Tang M and Zhang Y-C 2015
Chaos \textbf{25} 103102.

\bibitem{Miller2011}
Miller J C, Slim A C and Volz E M 2011 J. R. Soc. Interface.
\textbf{10} 1098.

\bibitem{Miller2013}
Miller J C and Volz E M 2013 PLoS ONE \textbf{8} e69162.

\bibitem{Wang2014}
Wang W, Tang M, Zhang H-F, Gao H, Do Y
and Liu Z-H 2014 Phys. Rev. E \textbf{90} 042803.

\bibitem{Karra2010}
Karrer B and Newman M E J 2010 Phys. Rev. E \textbf{82} 016101.

\bibitem{Newman2001}
Newman M E J, Strogatz S H and Watts D J 2001 Phys. Rev. E
\textbf{64} 026118.

\bibitem{Strogatz1994}
Strogatz S H 1994 \emph{Nonlinear dynamics and chaos: with applications to physics,
biology, chemistry and engineering} (Westview, Boulder, CO).

\bibitem{Dorogovtsev2008}
Dorogovtsev S N, Goltsev A V, and Mendes J F F 2008
Rev. Mod. Phys. \textbf{80} 1275.

\bibitem{Chen2014}
Chen W, Schr\"{o}der M, D'Souza M R 2014 Phys. Rev. Lett.
\textbf{112} 155701.

\bibitem{Radicchi2014}
Radicchi F 2015 Phys. Rev. E \textbf{91} 010801(R).

\bibitem{Ferreira2012}
Ferreira S C, Castellano C and Pastor-Satorras R 2015
Phys. Rev. E \textbf{86} 041125.

\bibitem{Shu2014}
Shu P, Wang W and Tang M 2015 Chaos \textbf{25} 063104.

\bibitem{Holme2003}
Holme P, Kim B J, Yoon C N, and Han S K.
2002 Phys. Rev. E \textbf{65} 056109.

\end{thebibliography}
\end{document}